\documentclass[twocolumn,prb,aps,amsfonts,amssymb,amsmath]{revtex4}
\usepackage{graphicx}
\usepackage{hyperref}
\begin{document}

%------------------------------------------------------
\title{Electron phase shift at the zero-bias anomaly of quantum point contacts}
%------------------------------------------------------

\author{B. Brun,$^{1,2}$ F. Martins,$^3$ S. Faniel,$^3$ B. Hackens,$^3$ A. Cavanna,$^4$ C. Ulysse,$^4$ A. Ouerghi,$^4$ U. Gennser,$^4$ \\ D. Mailly,$^4$ P. Simon,$^5$ S. Huant,$^{1,2}$ V. Bayot,$^{1,3}$ M. Sanquer,$^{1,6}$ and H. Sellier$^{1,2}$}

\email{hermann.sellier@neel.cnrs.fr}

\affiliation{
$^1$Universit\'e Grenoble Alpes, F-38000 Grenoble, France \\
$^2$CNRS, Institut NEEL, F-38042 Grenoble, France \\
$^3$IMCN/NAPS, Universit\'e catholique de Louvain, B-1348 Louvain-la-Neuve, Belgium \\
$^4$CNRS, Laboratoire de Photonique et de Nanostructures, UPR20, F-91460 Marcoussis, France \\
$^5$Laboratoire de Physique des Solides, Universit\'e Paris-Sud, F-91405 Orsay, France \\
$^6$CEA, INAC-SPSMS, F-38054 Grenoble, France
}

%\date{\today}

\begin{abstract}

The Kondo effect is the many-body screening of a local spin by a cloud of electrons at very low temperature. It has been proposed as an explanation of the zero-bias anomaly in quantum point contacts where interactions drive a spontaneous charge localization. However, the Kondo origin of this anomaly remains under debate, and additional experimental evidence is necessary. Here we report on the first phase-sensitive measurement of the zero-bias anomaly in quantum point contacts using a scanning gate microscope to create an electronic interferometer. We observe an abrupt shift of the interference fringes by half a period in the bias range of the zero-bias anomaly, a behavior which cannot be reproduced by single-particle models. We instead relate it to the phase shift experienced by electrons scattering off a Kondo system. Our experiment therefore provides new evidence of this many-body effect in quantum point contacts.

\end{abstract}

\maketitle

%------------------------------------------------------
%\section*{Introduction}
%------------------------------------------------------

Quantum point contacts \cite{wharam-88-jpc,vanwees-88-prl} (QPCs) are small constrictions in high-mobility two-dimensional electron gases (2DEGs) controlled by a metallic split gate at the surface of a semiconductor heterostructure. Despite their apparent simplicity, they reveal complex many-body phenomena which defy our understanding. When these quasi-one-dimensional ballistic channels are sufficiently open, electrons are perfectly transmitted via each available transverse mode \cite{buttiker-90-prb}, and the conductance is quantized in units of the conductance quantum $2e^2/h$. Below the first conductance plateau however, this single-particle picture fails due to the increasing importance of many-body effects. An additional shoulder shows up in the linear conductance curve around $0.7\times 2e^2/h$, called the 0.7 anomaly \cite{thomas-96-prl}, and a narrow peak of enhanced conductance appears around zero bias in the non-linear conductance curves at low enough temperature, called the zero-bias anomaly \cite{cronenwett-02-prl} (ZBA). The peak behavior versus temperature and magnetic field was shown to share strong similarities with the Kondo effect in quantum dots \cite{goldhabergordon-98-nat,cronenwett-98-sci} (QDs), i.e. the many-body screening of a local spin by conduction electrons below a characteristic temperature \cite{kondo-64-ptp,glazman-88-jetp,ng-88-prl}. However, deviations of the ZBA from the established Kondo effect have been reported \cite{cronenwett-02-prl,sfigakis-08-prl,sarkozy-09-prb,ren-10-prb}, and the occurrence of this effect in QPCs remains a debated issue \cite{micolich-11-jpcm,bauer-13-nat,heyder-15-prb}.

Because of enhanced electron interactions at low density, a spontaneous charge localization is predicted in QPCs below the first plateau \cite{rejec-06-nat,guclu-09-prb}, showing similarities with the one-dimensional Wigner crystallization \cite{wigner-34-pr,matveev-04-prl}. This phenomenon is supported by two recent experiments where localized states with even and odd numbers of charges have been observed \cite{iqbal-13-nat,brun-14-ncom}. The development of a Kondo effect is therefore expected at very low temperature, but its specific properties for a self-consistently localized state have not been calculated yet, due to the complexity of the problem. In this unsettled situation, the ZBA remains the subject of intensive investigations, and any new information pointing to a Kondo origin is important.

\begin{figure}[b]
\begin{center}
\includegraphics[width=7.5cm]{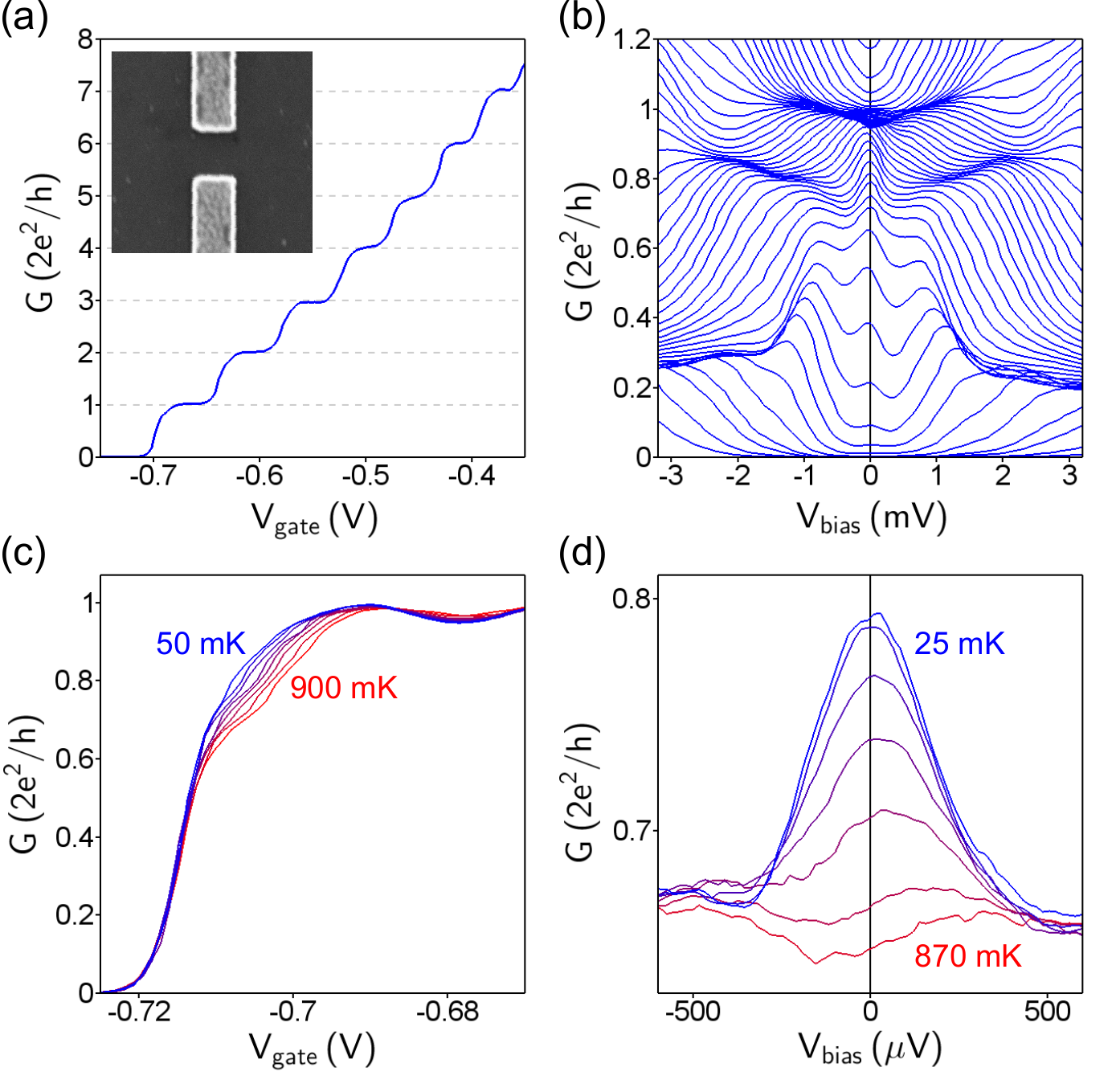}
\caption{(a) QPC conductance versus gate voltage at 30~mK (different cool down than other figures). Inset: image of the metallic split gate. (b) Differential conductance versus source-drain bias at 25~mK and different gate voltages. (c) Temperature dependence of the 0.7 anomaly from 50 to 900~mK. (d) Temperature dependence of the ZBA from 25 to 870~mK.}
\label{figure1}
\end{center}
\end{figure}

Here we use a scanning gate microscope \cite{topinka-01-nat} (SGM) to create a Fabry-P\'erot (FP) cavity between the QPC and the tip \cite{jura-09-prb,freyn-08-prl}, and measure the phase of the electron wave function scattered by the QPC in the ZBA regime. Phase-sensitive experiments indeed provide unique information on quantum phenomena and, in the case of QPCs, will help us to clarify the microscopic origin of the ZBA. Recently, a phase measurement on a QPC has been reported \cite{kobayashi-13-arxiv}, but no significant deviation from the single-particle prediction has been found \cite{lackenby-14-prb}. In the past, the transmission phase of QDs in the Kondo regime was measured by embedding them in Aharonov-Bohm (AB) rings \cite{ji-00-sci,zaffalon-08-prl,takada-14-prl}. Here we measure instead the reflection phase of the system and observe a phase shift by $\pi$ of the interference fringes in the bias voltage range of the ZBA. This shift occurs via two phase jumps, and disappears with gate voltage and temperature in the same way as the ZBA. Calculations of the reflection phase for a single-particle resonant level give a smooth shift across the resonance \cite{buks-96-prl}, in strong contrast with the two phase jumps observed in our experiment, thereby indicating a many-body origin. The observed behavior shows characteristic signatures of the Kondo effect, where the transmission phase at the Fermi energy is locked at $\pi/2$ in the Kondo valleys \cite{glazman-88-jetp}, and where a ``sharp Kondo double phase lapse'' is predicted as a function of source-drain bias \cite{gerland-00-prl}. We therefore attribute the observed phase shift to the Kondo effect, thus providing new evidence of this effect as the origin of the zero-bias anomaly in QPCs.

\begin{figure}[b]
\begin{center}
\includegraphics[width=8.5cm]{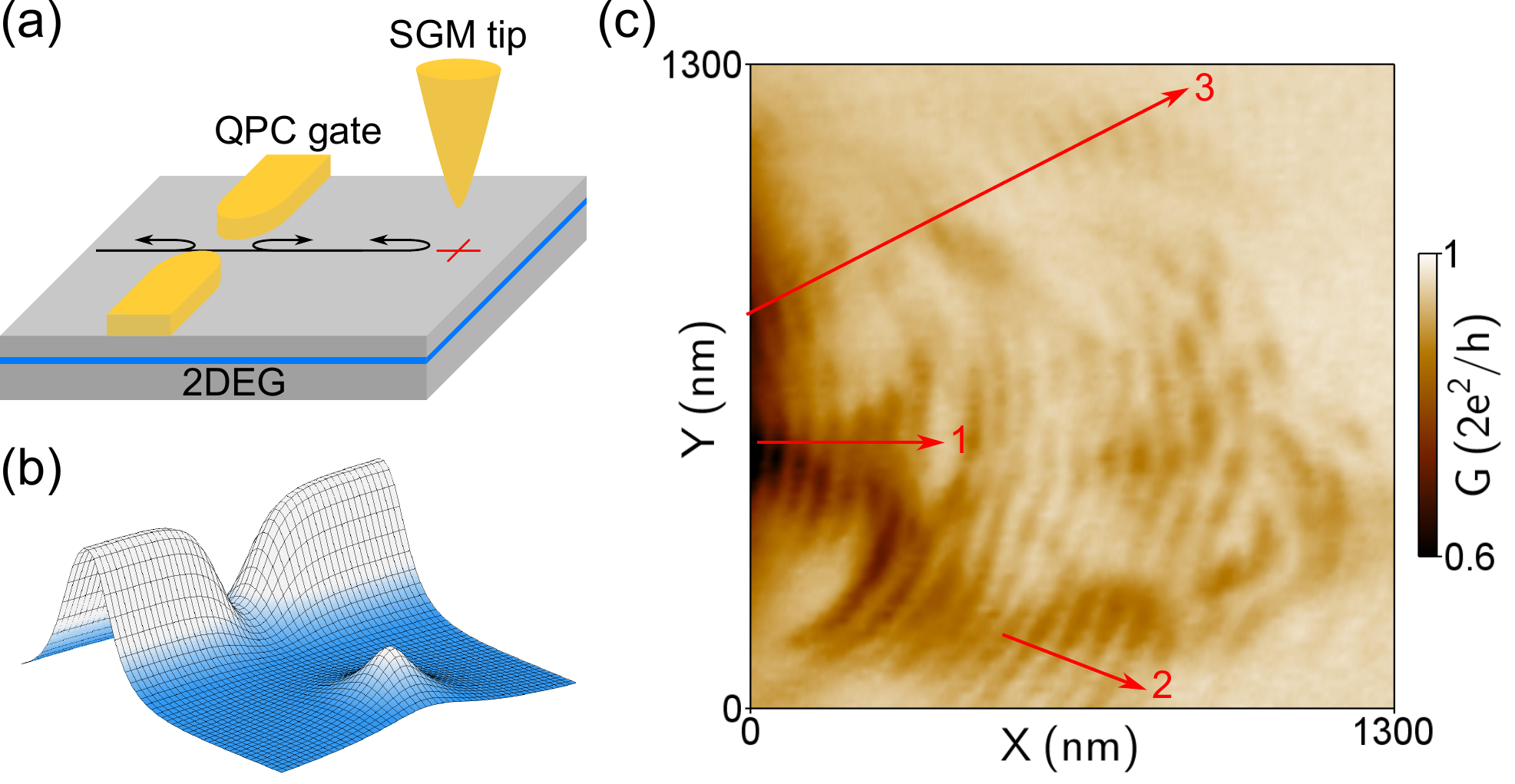}
\caption{(a) Principle of scanning gate interferometry. (b) Potential landscape created by the split gate and the tip. (c) SGM image of the conductance at 25~mK when the QPC is tuned to the first conductance plateau. The QPC center is located at the coordinates ($-500$~nm, 650~nm).}
\label{figure2}
\end{center}
\end{figure}

\begin{figure}[t]
\begin{center}
\includegraphics[width=8.5cm]{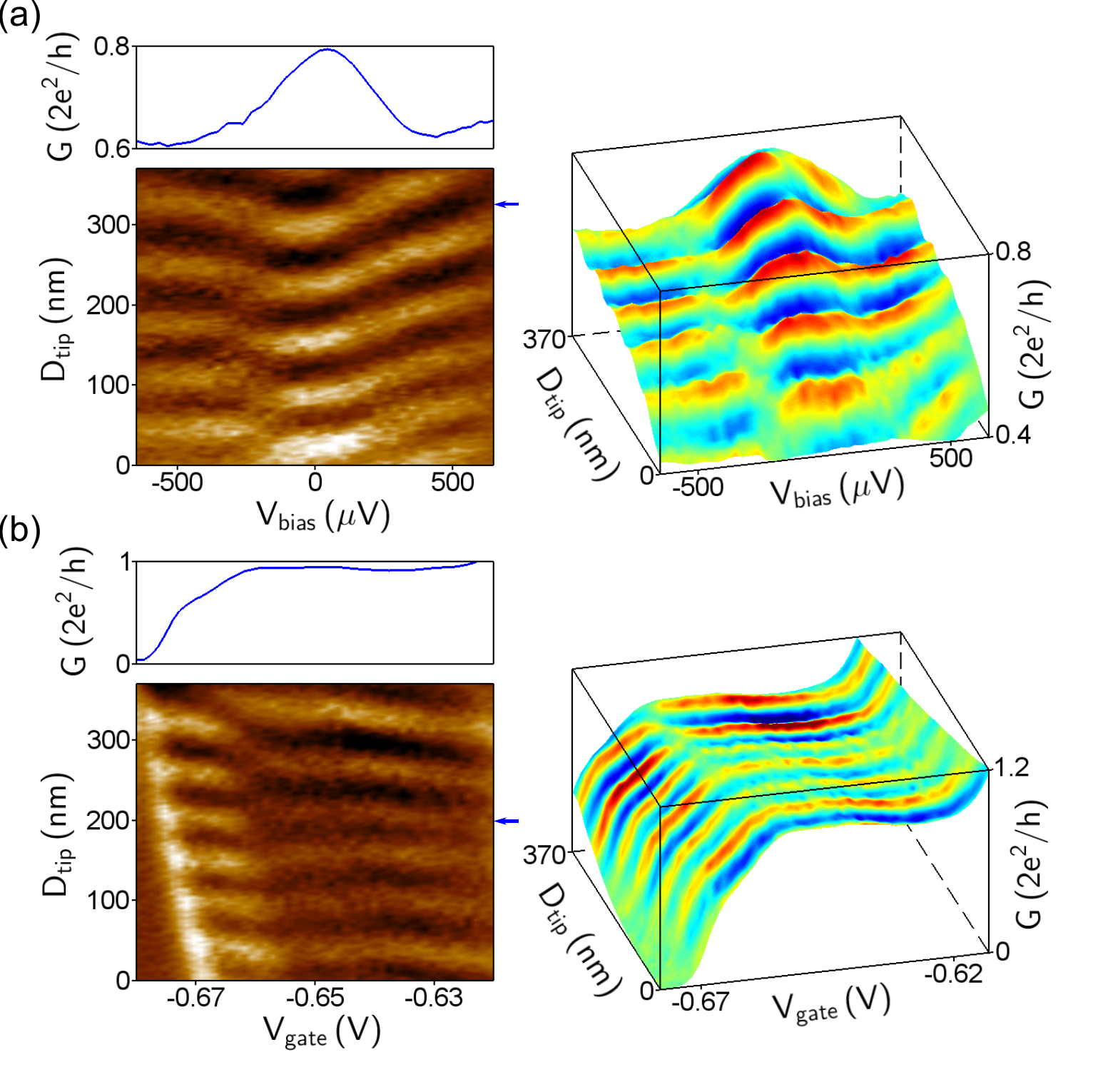}
\caption{(a) Interference fringes along line 1 (Fig.~\ref{figure2}(c)) versus source-drain bias at $-0.67$~V gate voltage in (b). The conductance is differentiated with respect to tip position. Top panel: conductance curve for the tip position indicated by the arrow. Right panel: 3D plot showing the position of the phase shift at the bottom of the zero-bias peak. (b) Interference fringes along the same line versus gate voltage (at zero bias). The pinch-off voltage is shifted by 40~mV in the presence of the polarized tip with respect to Fig.~\ref{figure1}(c). Top panel: conductance curve for the tip position indicated by the arrow. Right panel: 3D plot showing the position of the phase shift at the border of the plateau.}
\label{figure3}
\end{center}
\end{figure}

\begin{figure}[b]
\begin{center}
\includegraphics[width=8.5cm]{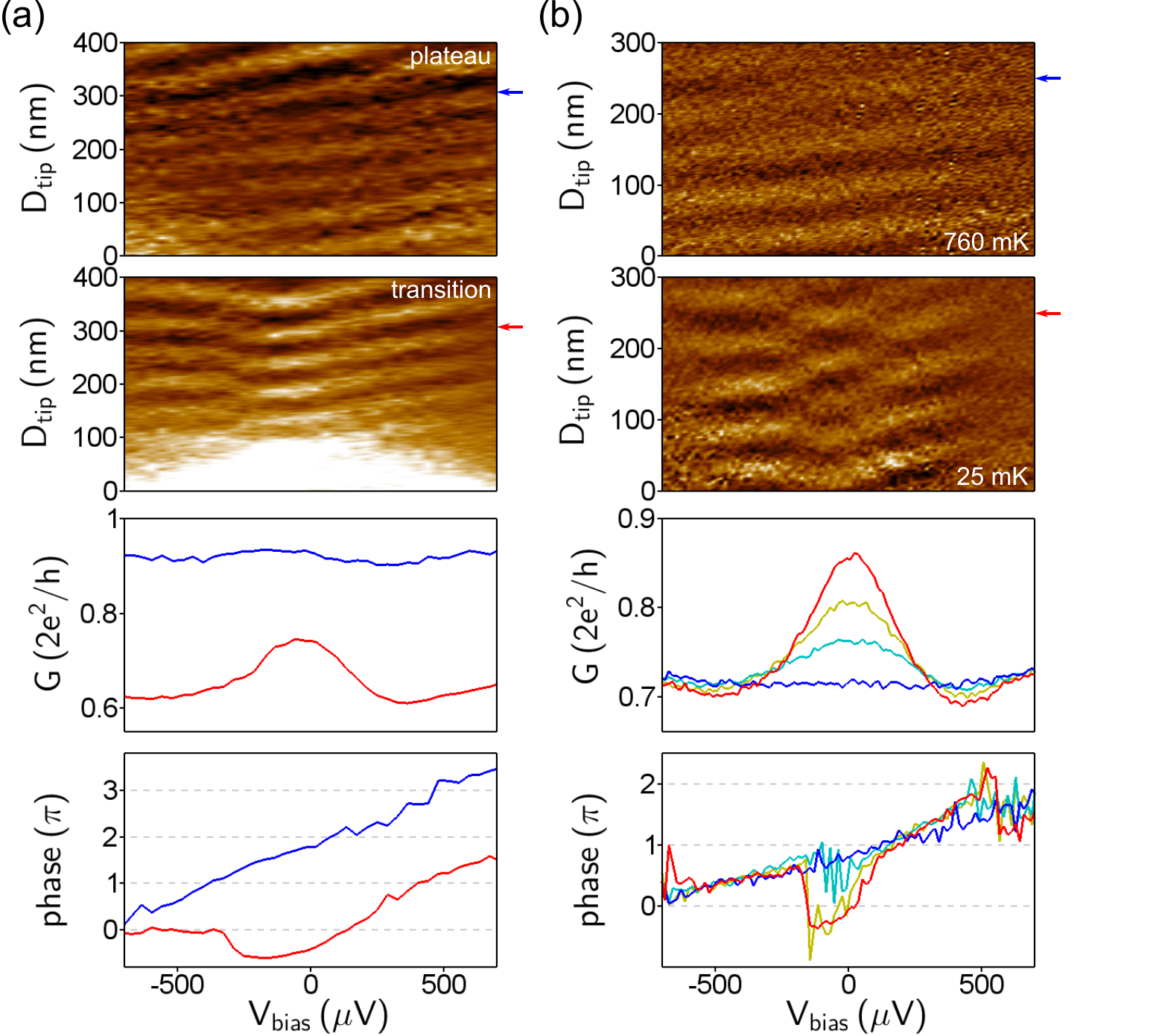}
\caption{(a) Interference fringes along line 1 on the plateau (first panel) and below the plateau (second panel) at respectively $-0.65$~V and $-0.67$~V gate voltages in Fig.~\ref{figure3}(b). Third panel: conductance curve at the tip position of the arrow, below the plateau (red curve) and on the plateau (blue curve). Bottom panel: the phase of the fringes exhibits a shift in the bias range of the ZBA (red curve) and evolves linearly on the plateau (blue curve). (b) Interference fringes along line 2 at a gate voltage below the plateau at 760~mK (first panel) and 25~mK (second panel). Third panel: conductance curve at the tip position of the arrow, at 25, 240, 440, and 760~mK from top to bottom. Bottom panel: phase of the fringes at the same temperatures.}
\label{figure4}
\end{center}
\end{figure}

%------------------------------------------------------
%\section*{Results}
%------------------------------------------------------

The QPC is defined in a GaAs/AlGaAs heterostructure by the 270~nm long and 300~nm wide gap of a Ti/Au split gate (inset of Fig.~\ref{figure1}(a)). The 2DEG located 105~nm below the surface has a $2.5 \times 10^{11}$ cm$^{-2}$ electron density and a $1.0 \times 10^6$ cm$^2$V$^{-1}$s$^{-1}$ electron mobility. The device is fixed to the mixing chamber of a dilution fridge in front of a cryogenic scanning probe microscope \cite{hackens-10-ncom,martins-13-sr} and cooled down to a base temperature of 25~mK at zero gate voltage. The four-probe differential conductance is measured by a lock-in technique using a 10~$\mu$V excitation. A series resistance of 1600~$\Omega$ is subtracted from all data in order to have the first conductance plateau at $2e^2/h$.

At the base temperature (25~mK), the linear conductance shows quantized plateaus and smooth transitions versus gate voltage (Fig.~\ref{figure1}(a)), while at higher temperatures, the conductance exhibits the well-known 0.7 anomaly \cite{thomas-96-prl} below the first plateau (Fig.~\ref{figure1}(c)). The non-linear differential conductance versus source-drain bias shows a narrow peak at zero bias (Fig.~\ref{figure1}(b)), the so-called ZBA \cite{cronenwett-02-prl}, which vanishes rapidly at higher temperatures (Fig.~\ref{figure1}(d)). The temperature dependence of the peak height can be rescaled on the universal scaling law of Kondo QDs \cite{goldhabergordon-98-prl,vanderwiel-00-sci,nygard-00-nat} with a single scaling parameter $T_K$, called Kondo temperature, for all gate voltages (Fig.~S1 in the Supplemental Material \cite{supplemental}). 

We now investigate the scattering phase of the QPC in the ZBA regime at very low temperature (25~mK) using a SGM-based interferometry experiment (Fig.~\ref{figure2}(a)). The SGM tip is scanned above the 2DEG at finite distance from the QPC, with a tip voltage of $-6$~V and a tip-to-surface height of 30~nm, chosen such as to locally deplete the 2DEG (Fig.~\ref{figure2}(b)). Electrons propagating out of the QPC are scattered by this tip-induced perturbation and partially reflected towards the QPC. Interference fringes show up in the SGM images (Fig.~\ref{figure2}(c)) due to the coherent superposition of waves reflected by the QPC and the tip, forming together a FP cavity. To probe the scattering phase at the ZBA, the tip is scanned along individual lines where regular fringe patterns are observed (red lines). In the ZBA region below the first plateau, a shift of the interference fringes appears around zero source-drain bias, with abrupt jumps on each side of the ZBA (Fig.~\ref{figure3}(a)). When the fringes are recorded while sweeping the gate voltage (Fig.~\ref{figure3}(b)), a similar shift is observed when the conductance drops below the first plateau, i.e. when the QPC enters the ZBA region. This phase shift reveals the non-trivial scattering phase of the ZBA and constitutes a new experimental signature of this many-body effect.

The phase of the interference fringes in various situations is extracted in Fig.~\ref{figure4} from a Fourier transform performed along the scan axis. When the QPC is tuned to the first plateau (Fig.~\ref{figure4}(a), top panel), the fringes evolve monotonically with source-drain bias due to a change in wavelength for electrons injected at higher energy \cite{gorini-14-prb}, and the extracted phase is linear (blue curve, bottom panel). Below the first plateau (second panel), the fringes exhibit a sharp phase jump at negative bias and a smooth one at positive bias, also visible on the extracted phase (red curve, bottom panel). These phase jumps occur when the conductance increases above the background to build the zero-bias peak (red curve, third panel). Figure~S2 in the Supplemental Material \cite{supplemental} presents additional data.

In order to measure the zero-bias phase shift, it is necessary to have a reference phase at the same gate voltage for a situation without the ZBA. This can be obtained by recording the interference pattern at different temperatures and fixed gate voltage (Fig.~\ref{figure4}(b)). At a temperature where the ZBA has disappeared (top panel), the phase evolves linearly (blue curve, bottom panel), whereas at the lowest temperature where the ZBA is at maximum (second panel), the phase shows two jumps with a shift of about $\pi$ (red curve, bottom panel). At intermediate temperatures, the phase jumps remain at the same bias voltages, but the shift disappears progressively, in a non-uniform way, explaining larger fluctuations in the extracted phase (Fig.~S3 \cite{supplemental}).

\begin{figure}[t]
\begin{center}
\includegraphics[width=8cm]{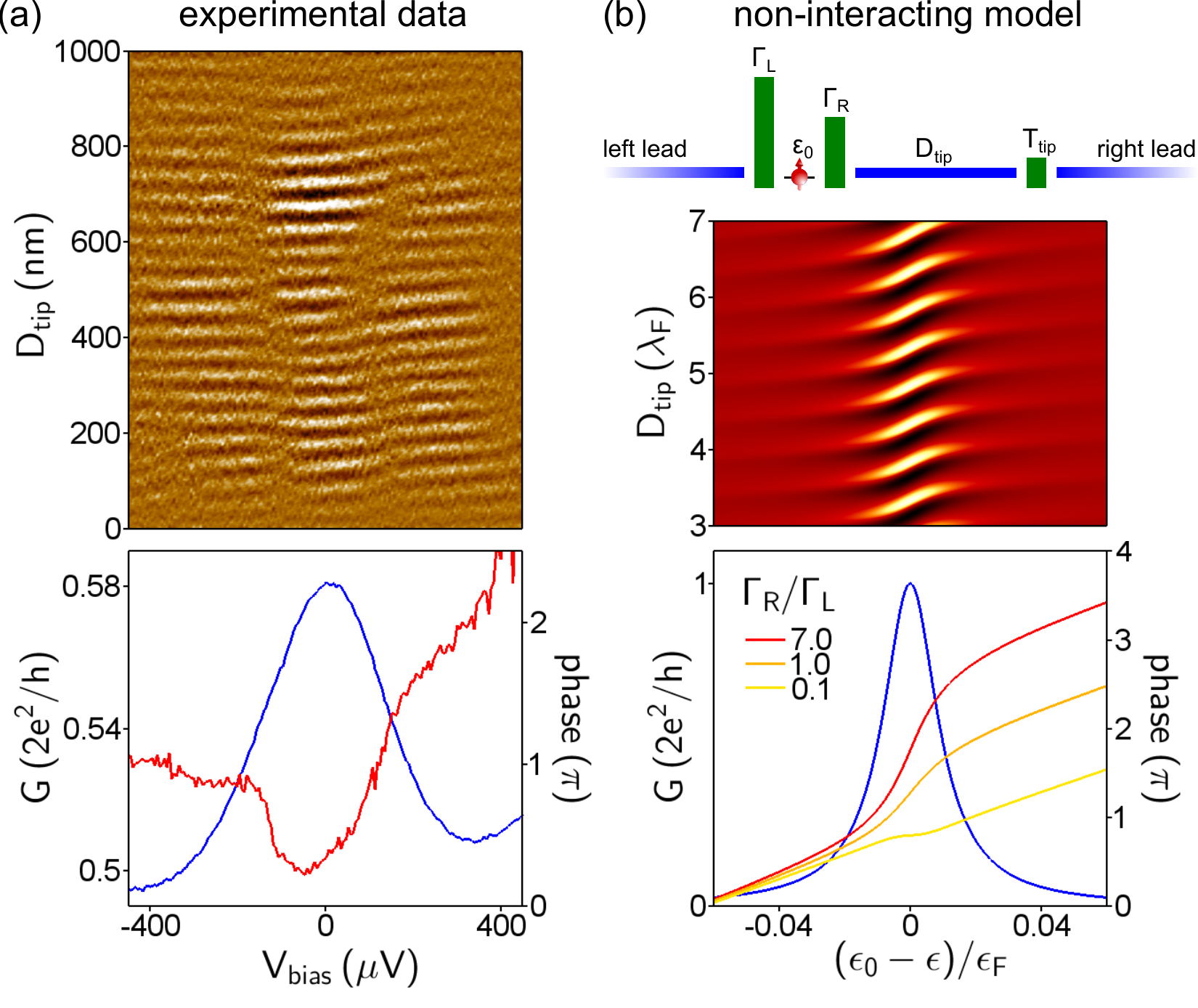}
\caption{(a) Interference fringes when the tip is scanned along line 3. Bottom panel: average conductance curve (blue) and phase of the fringes (red) showing a shift by $\pi$ within the ZBA. (b) Model of the SGM-based interferometry experiment where the QPC is represented by an asymmetric QD. Bottom panel: conductance in the symmetric case (blue) and phase of the interference fringes versus energy, for different asymmetries of the tunneling rates $\Gamma_L$ and $\Gamma_R$ (red to yellow). Central panel: tip-induced interference fringes for the asymmetry of the red curve.}
\label{figure5}
\end{center}
\end{figure}

A better accuracy on the phase determination can be obtained by choosing a longer scanning line with more interference fringes, but the difficulty is to find such a long line where the ZBA remains relatively constant along the entire scan. Indeed, as reported in Ref.~\cite{brun-14-ncom}, the ZBA splits up into finite bias peaks due to a periodic change of the localized state occupancy with tip distance, and this limits the available scan lengths. However, when the tip is scanned along the red line 3, the interference fringes are regularly spaced (Fig.~\ref{figure5}(a), top panel) and the ZBA is only slightly disturbed. The phase (bottom panel) shows an abrupt jump at negative bias and a smoother change at positive bias, with a zero-bias shift close to $\pi$. The phase shift is also observed versus gate voltage along this scanning line (Fig.~S4 \cite{supplemental}).

% ------------------------------------------------------
%\section*{Discussion}
%------------------------------------------------------

In our experiment, the sensitivity of the interference pattern to the ZBA, which is an intrinsic QPC property, demonstrates that the QPC is part of the interferometric cavity. The QPC represents one of the cavity mirrors, as also realized in Refs.~\cite{jura-09-prb,kozikov-13-njp}, but in contrast to experiments where interference was attributed to impurities in the 2DEG \cite{topinka-01-nat,jura-07-natp}. This situation is consistent with the fact that the interference fringes are observed within the thermal length $L_T=\hbar v_F/k_B T$ which is 1.5~$\mu$m at 1~K and much more below (Fig.~S5 \cite{supplemental}). In addition, the zero-bias phase shift is observed for all scanning lines that have been investigated, showing that it really corresponds to the scattering phase of the QPC, and does not result from specific scatterers in the 2DEG region between the QPC and the tip. It has also been observed in a second device (Fig.~S6 \cite{supplemental}).

For quantitative analysis, it is important to note that our SGM experiment realizes a FP cavity \cite{rossler-15-prl} and therefore probes the reflection phase of the QPC. This situation differs from previous experiments on QDs using AB rings \cite{schuster-97-nat} which probe the transmission phase of the embedded device. In the case of a single-particle resonant level in a QD, the transmission phase presents a smooth shift by $\pi$ across the resonance \cite{schuster-97-nat}, while the reflection phase of an asymmetric QD presents a shift by 0 or $2\pi$ depending on which side the highest barrier is located \cite{buks-96-prl}. The reflection phase measured in our SGM experiment is therefore between zero and twice the transmission phase of the QPC and should be interpreted carefully.

The spontaneous charge localization in QPCs results from the formation of self-consistent barriers along the channel \cite{rejec-06-nat,guclu-09-prb}. The QPC can thus be modeled by a small QD with two asymmetric barriers on top of the main potential barrier controlled by the gate \cite{ren-10-prb}. The phase of tip-induced interference fringes has been calculated for non-interacting electrons using this simple model (Fig.~\ref{figure5}(b) and Fig.~S7 \cite{supplemental}). For all barrier asymmetries, the calculated phase exhibits a single smooth shift across the resonance (Fig.~\ref{figure5}(b), bottom panel), in contrast with the experimental behavior showing phase jumps on both sides of the resonance (Fig.~\ref{figure5}(a)). This difference indicates that the observed phase shift does not result from scattering on a localized state. The spontaneously localized states are indeed expected at larger energy and to survive up to much higher temperatures~\cite{yoon-07-prl}. Here, we are dealing with a low-energy phenomenon, that we attribute to the screening of the localized states by the Kondo effect at very low temperature \cite{kondo-64-ptp,glazman-88-jetp,ng-88-prl}. This screening produces a narrow resonance in the density of states at the Fermi level and gives rise to a conductance peak at zero bias \cite{hershfield-91-prl}.

Below the Kondo temperature, the transmission phase of a symmetric QD equals $\pi/2$ in the gate voltage range of a Kondo valley \cite{ji-00-sci,zaffalon-08-prl,takada-14-prl}, and the conductance reaches $2e^2/h$ \cite{vanderwiel-00-sci,kretinin-11-prb}. The phase shift observed in our experiment at zero bias may correspond to this Kondo scattering phase, but in the reflection coefficient, which can be twice the value of the transmission coefficient \cite{buks-96-prl}. This situation arises if the smallest barrier is located on the cavity side, which is likely to occur since the main gate-controlled barrier induces this asymmetry on the self-consistent confinement potential \cite{ren-10-prb}. A phase shift by $\pi$ is therefore expected at zero bias, which is close to the value found experimentally.

At finite bias voltage, the Kondo phase shift has been calculated in Ref.~\cite{gerland-00-prl} for a QD in equilibrium (Fig.~S8 \cite{supplemental}). It exhibits three switches from 0 to $\pi$ corresponding to the transmission through the single-particle level (first and second electrons) and through the Kondo resonance (always centered at zero bias). A ``sharp Kondo double phase lapse'' has been predicted around the Kondo peak at low enough temperature \cite{gerland-00-prl}, and the double phase jump seen in our experiment around the ZBA may correspond to such an effect. Phase lapses by $\pi$ are usually observed \textit{versus gate voltage} between the successive charge states of QDs in the Coulomb blockade regime, and explained by the coupling of the different orbitals to the leads \cite{oreg-97-prb,karrasch-07-prl,hecht-09-prb}. But to our knowledge, phase lapses \textit{versus source-drain bias} have not been reported before. In addition, decoherence of the Kondo correlations at finite bias voltage \cite{hershfield-91-prl,meir-93-prl} is also an effect that should be considered, but no theoretical prediction of the Kondo phase shift out of equilibrium exists at the moment. We expect our experiment to stimulate theoretical works in this direction.

%------------------------------------------------------
%\section*{Conclusion}
%------------------------------------------------------

To conclude, we performed the first phase-sensitive measurements on the QPC conductance anomalies using scanning gate interferometry. Whenever the ZBA is present, a phase shift of the interference fringes is observed around zero bias, and we interpret it as the Kondo phase shift experienced by electrons at the Fermi level. In addition, the two phase jumps around the conductance peak may correspond to the predicted phase lapses around the Kondo resonance. These results reinforce our understanding of the ZBA in terms of a Kondo effect on spontaneously localized states.

%------------------------------------------------------
%\section*{Acknowledgements}
%------------------------------------------------------

We thank J.-L. Pichard, D. Weinmann, X. Waintal, H. Baranger, M. Lavagna, S. Florens, and T. Meunier for discussions. This work was supported by the French Agence Nationale de la Recherche (``ITEM-exp'' project), by FRFC Grant No. 2.4503.12, and by FRS-FNRS Grants No. 1.5.044.07.F and No. J.0067.13. F.M. and B.H. acknowledge support from the Belgian FRS-FNRS, S.F. received support from the FSR at UCL, and V.B. acknowledges the award of a ``chair of excellence'' by the Nanosciences foundation in Grenoble.

%------------------------------------------------------
%\section*{References}
%------------------------------------------------------

%------------------------------------------------------

%------------------------------------------------------
% Supplementary Information
%------------------------------------------------------

\clearpage
\onecolumngrid
\addtolength{\textheight}{2cm}
\setcounter{figure}{0}
\setcounter{section}{0}
\renewcommand{\thefigure}{S\arabic{figure}}

%------------------------------------------------------
\begin{center}

\textbf{\large Supplemental Material for \\ Electron phase shift at the zero-bias anomaly of quantum point contacts} \vspace{5mm}

B. Brun,$^{1,2}$ F. Martins,$^3$ S. Faniel,$^3$ B. Hackens,$^3$ A. Cavanna,$^4$ C. Ulysse,$^4$ A. Ouerghi,$^4$ U. Gennser,$^4$ \\ D. Mailly,$^4$ P. Simon,$^5$ S. Huant,$^{1,2}$ V. Bayot,$^{1,3}$ M. Sanquer,$^{1,6}$ and H. Sellier$^{1,2}$

\textit{
$^1$Universit\'e Grenoble Alpes, F-38000 Grenoble, France \\
$^2$CNRS, Institut NEEL, F-38042 Grenoble, France \\
$^3$IMCN/NAPS, Universit\'e catholique de Louvain, B-1348 Louvain-la-Neuve, Belgium \\
$^4$CNRS, Laboratoire de Photonique et de Nanostructures, UPR20, F-91460 Marcoussis, France \\
$^5$Laboratoire de Physique des Solides, Universit\'e Paris-Sud, F-91405 Orsay, France \\
$^6$CEA, INAC-SPSMS, F-38054 Grenoble, France
}

\end{center}
%------------------------------------------------------

%------------------------------------------------------
\section{Scaling analysis of the zero-bias conductance versus temperature}
%------------------------------------------------------

\begin{figure}[h]
\begin{center}
\includegraphics[width=18cm]{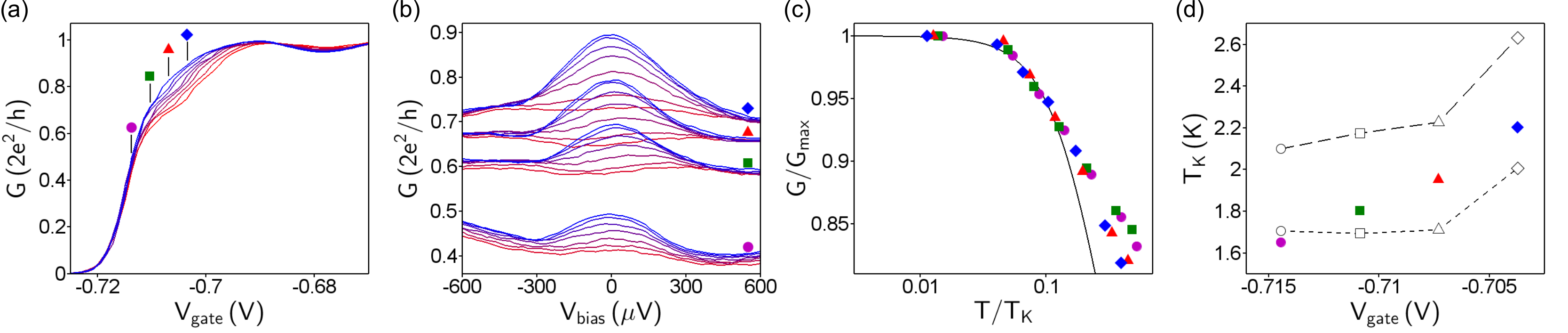}
\caption{(a) Conductance at zero bias versus gate voltage from 50~mK (blue) to 900~mK (red). (b) Differential conductance versus source-drain bias from 25~mK (blue) to 870~mK (red), at the four gate voltages indicated by the colored symbols in (a). Temperatures are 25, 90, 145, 230, 380, 650, 870~mK. (c) The conductance at zero-bias from (b) is normalized to the lowest temperature value and plotted versus a rescaled temperature $T/T_K$. The symbol color corresponds to the data taken from (b) at the gate voltage indicated in (a). The solid line indicates the universal Kondo behavior. (d) The Kondo temperatures $T_K$ used in the scaling analysis in (c) are plotted versus gate voltage as colored symbols. The graph also shows (open symbols) the characteristic temperatures $T_K^*=e\Delta V/2k_B$ given by the full width $\Delta V$ of the zero-bias peaks shown in (b) at 25~mK. Top symbols (connected by a dashed line) are obtained using the full width at 1/2 of the peak maximum (FWHM). Bottom symbols (connected by a dotted line) are obtained using the full width at 2/3 of the peak maximum (FW2/3M).}
\label{figure-S1}
\end{center}
\end{figure}

The temperature dependence of the linear conductance at zero bias is shown in Fig.~\ref{figure-S1}(a). The 0.7 anomaly observed at high temperature is related to the suppression of the conductance peak observed versus source-drain bias as shown in Fig.~\ref{figure-S1}(b). The zero-bias conductance $G$ at finite temperature, normalized to its value $G_{\rm max}$ at the lowest temperature, is plotted in Fig.~\ref{figure-S1}(c) as a function of the temperature $T$, rescaled by a parameter $T_{\rm K}$ called Kondo temperature. This parameter is chosen such that the data points follow the universal scaling law \cite{costi-94-jpcm} of Kondo quantum dots (QDs) which can be approximated by the phenomenological formula \cite{goldhabergordon-98-prl,vanderwiel-00-sci,nygard-00-nat}:
\begin{eqnarray*}
G(T/T_{\rm K})=G_{\rm max}(1+(2^{1/s}-1)(T/T_{\rm K})^2)^{-s}
\end{eqnarray*}
where $s$ is a fixed parameter close to 0.2 for $S=1/2$ Kondo QDs. The single scaling parameter $T_{\rm K}$ is shown in Fig.~\ref{figure-S1}(d) and depends on the gate voltage. Note that the conductance $G_{\rm max}$ at zero temperature is less than $2e^2/h$ and might be expressed as \cite{ng-88-prl}:
\begin{eqnarray*}
G_{\rm max}=(2e^2/h)4\Gamma_{\rm L}\Gamma_{\rm R}/(\Gamma_{\rm L}+\Gamma_{\rm R})^2
\end{eqnarray*}
using asymmetric tunneling rates $\Gamma_{\rm L,R}$ for the main gate-controlled QPC barrier and the weak self-consistent barrier resulting from Coulomb interactions \cite{ren-10-prb}.

In Kondo QDs \cite{vanderwiel-00-sci,nygard-00-nat} and in QPCs \cite{ren-10-prb,cronenwett-02-prl,sarkozy-09-prb}, the full width at half-maximum (FWHM) of the zero-bias peak is often assumed to equal $2k_{\rm B}T_{\rm K}/e$, but recent investigations \cite{pletyukhov-12-prl,kretinin-12-prb,klochan-13-prb} have shown that the full width at 2/3 of the maximum (FW2/3M) gives a better estimate of $T_{\rm K}$. In our QPC, the peak width (almost temperature independent) is plotted in Fig.~\ref{figure-S1}(d) as a function of gate voltage. The FW2/3M values (open symbols and dotted line) show a slightly better agreement with the $T_{\rm K}$ values from the scaling analysis than the FWHM values (open symbols and dashed line).

In the absence of a generally accepted theory of the Kondo effect in QPCs, we have also investigated the possibility of a $S=1$ Kondo effect, for which the parameter $s$ is close to 0.16 \cite{costi-09-prl}. However, the quality of the fit is not better, and the new set of Kondo temperatures is found to be 1.4 times larger, in worse agreement with the values of the peak width, in particular with the FW2/3M values. Since the $S=1/2$ Kondo model describes better the behavior of the ZBA, we have here an indication of the Kondo mechanism responsible for the ZBA in QPCs.

\newpage
%------------------------------------------------------
\section{Evolution of the phase shift with gate voltage (line 1)}
%------------------------------------------------------

\begin{figure}[h]
\begin{center}
\includegraphics[width=17cm]{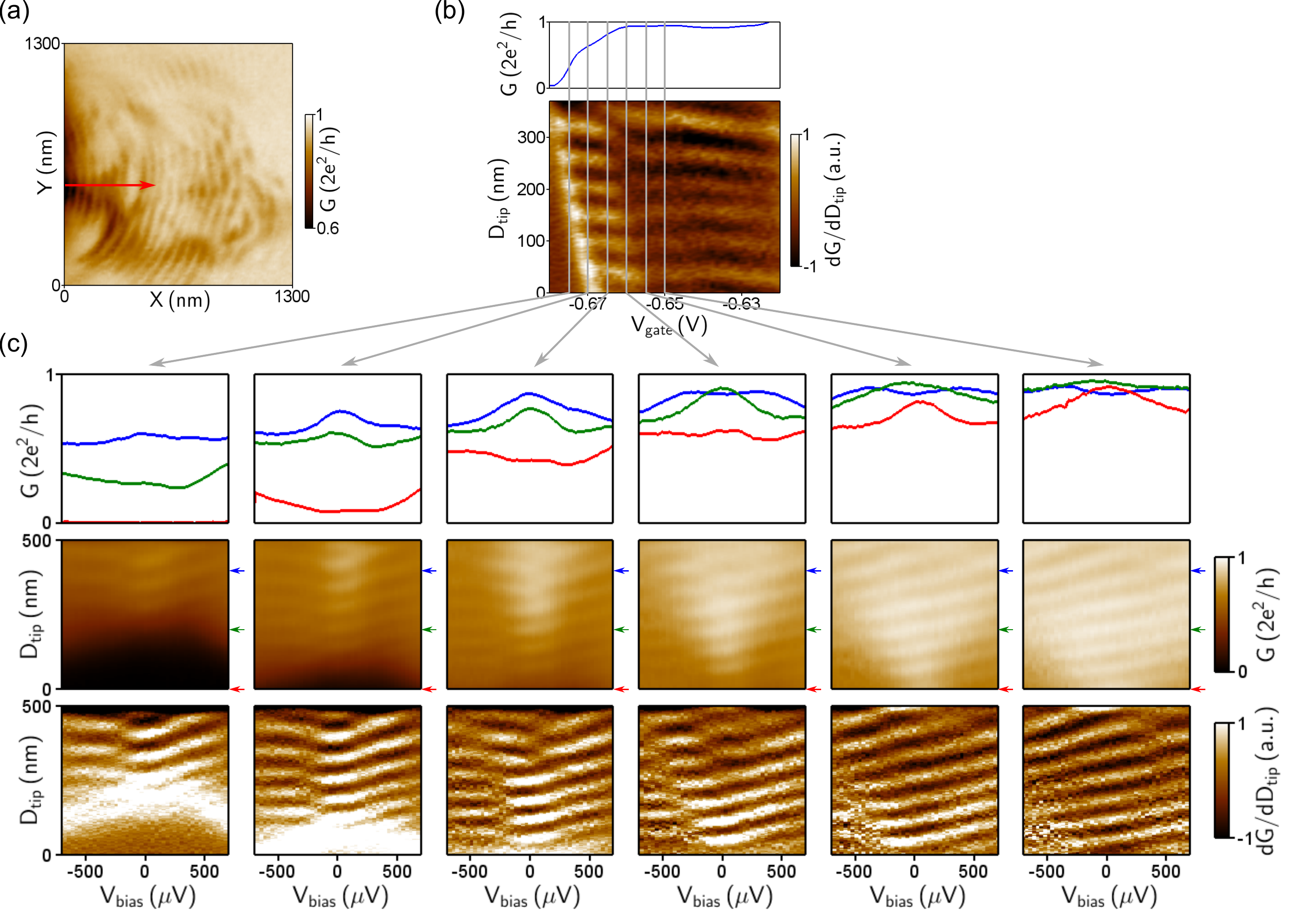}
\caption{(a) Same SGM image as in Fig.~2(c). The QPC is on the first conductance plateau. (b) Same data as in Fig.~3(b). The interference fringes at zero bias along the red line in (a) are plotted versus gate voltage. Top panel: conductance curve at 200~nm tip distance. (c) Interference fringes along the red line in (a) versus source-drain bias, for gate voltages every 5~mV from -0.675 to -0.650~V (from left to right). Central panels: raw conductance plot. Bottom panels: derivative of the conductance with respect to tip distance. Top panels: conductance curves at 0, 200, and 400~nm (from bottom to top). A phase shift with two phase jumps are visible around the ZBA when the conductance is below $0.8\times 2e^2/h$. Above this value, the ZBA splits up in two finite bias peaks, as reported in Ref.~\cite{brun-14-ncom}, and the interference fringes shows an additional phase jump at zero-bias.}
\label{figure-S2}
\end{center}
\end{figure}

\newpage
%------------------------------------------------------
\section{Evolution of the phase shift with temperature (line 2)}
%------------------------------------------------------

\begin{figure}[h]
\begin{center}
\includegraphics[width=15cm]{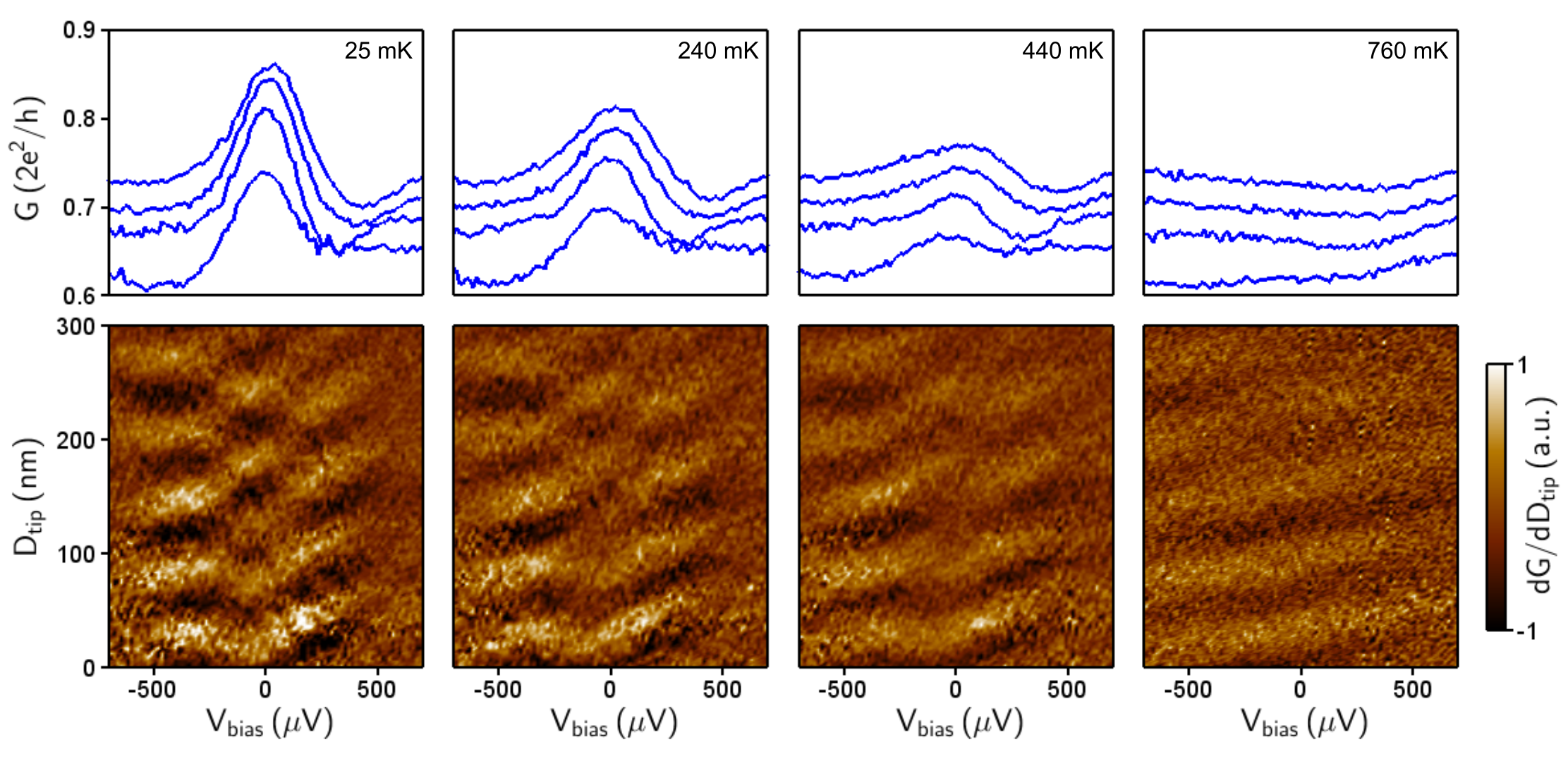}
\caption{Interference fringes versus source-drain bias for different temperatures. This figure shows the data at 240 and 440~mK mentioned in Fig.~4(b), together with the data at 25 and 760~mK already shown in Fig.~4(b). The gate voltage is -0.68~V and the tip is scanned along line 2 in Fig.~2(c). Top panels: conductance curves at 0, 100, 200, and 300~nm (from bottom to top). The phase shift disappears with increasing temperature in the same way as the ZBA.}
\label{figure-S3}
\end{center}
\end{figure}

%------------------------------------------------------
\section{Phase shift versus bias voltage and gate voltage (line 3)}
%------------------------------------------------------

\begin{figure}[h]
\begin{center}
\includegraphics[width=11cm]{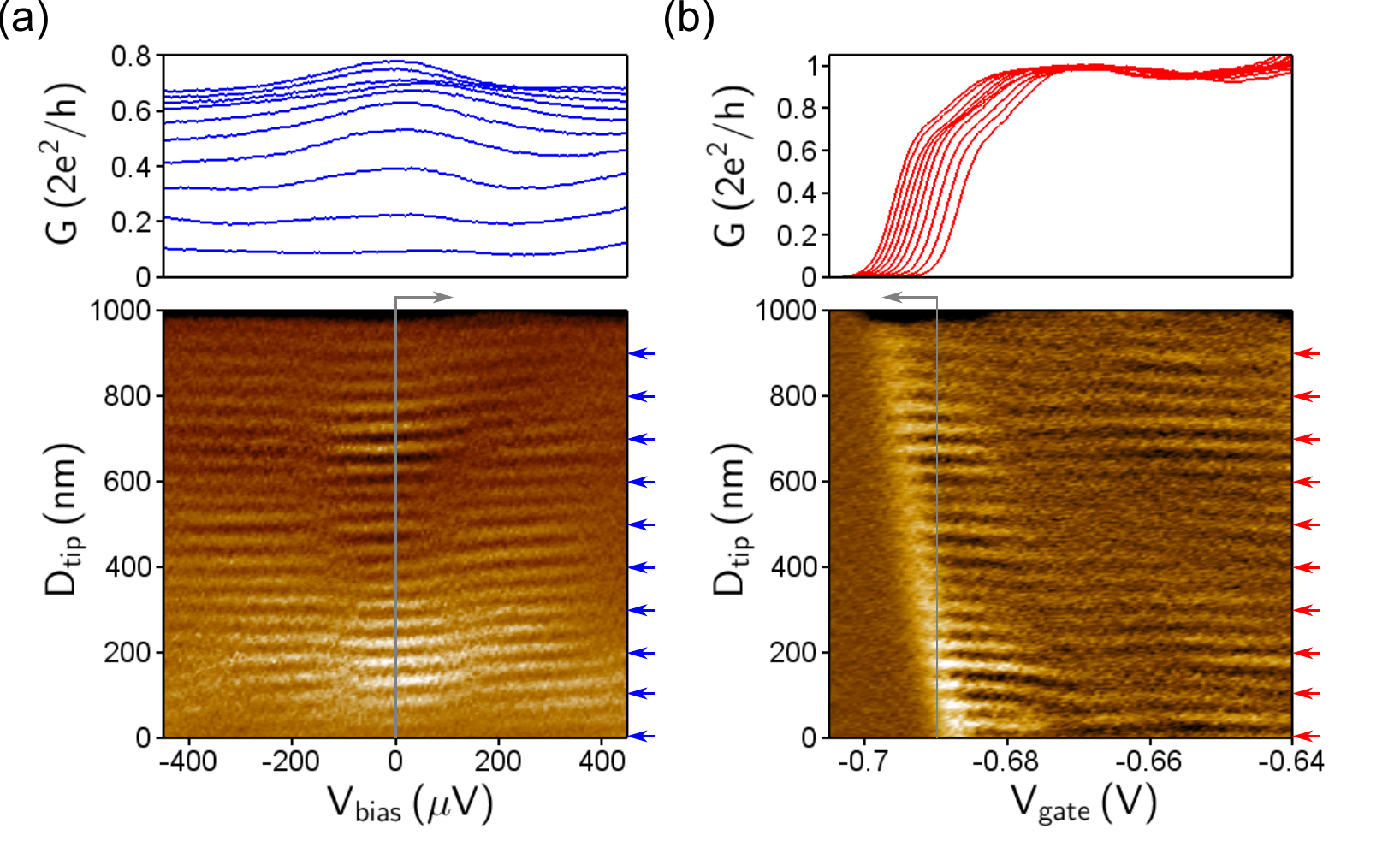}
\caption{Interference fringes when the tip is scanned along line 3 in Fig.~2(c). (a) Interference fringes versus source-drain bias at -0.69~V gate voltage (same data as in Fig.~5(a)). Top panel: conductance curves every 100~nm from 0 to 900~nm (from bottom to top). (b) Interference fringes versus gate voltage at zero bias. Top panel: conductance curves every 100~nm from 0 to 900~nm (from right to left). The fringes show a phase shift at the border of the conductance plateau, similar to that shown in Fig.~3(b) which was measured along line 1.}
\label{figure-S4}
\end{center}
\end{figure}

\newpage
%------------------------------------------------------
\section{Temperature dependence of the fringes visibility}
%------------------------------------------------------

\begin{figure}[h]
\begin{center}
\includegraphics[width=11cm]{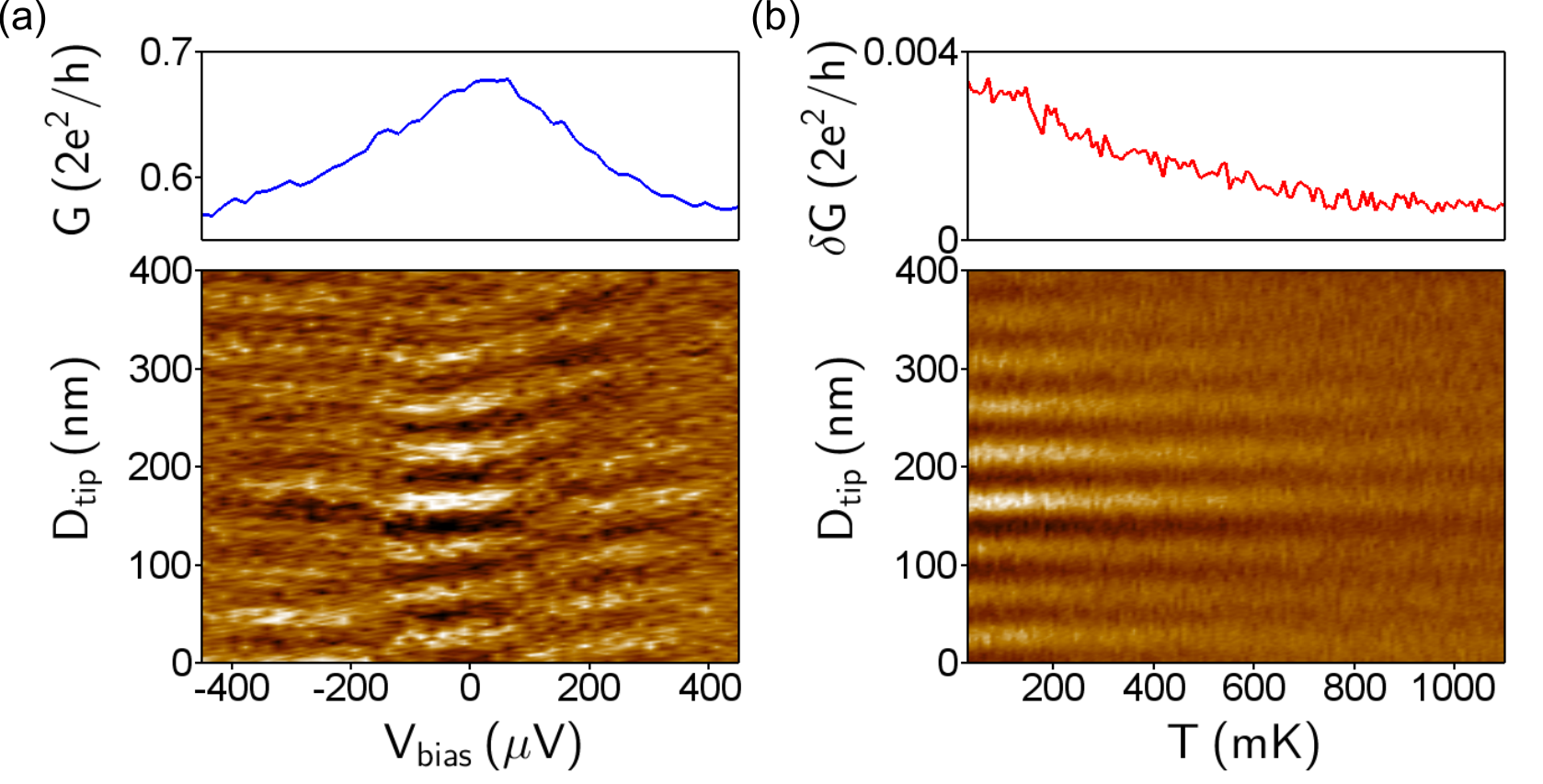}
\caption{(a) Interference fringes versus source-drain bias (below the first conductance plateau) when the tip is scanned along a line close to the end of line 3 in Fig.~2(c). Top panel: conductance curve at 0~nm. (b) Interference fringes at zero-bias along the same scanning line, measured as a function of temperature. Top panel: evolution of the fringes visibility, showing an exponential decay on a characteristic temperature of about 800~mK. Above this temperature, the thermal length $L_T=\hbar v_F/k_B T$ becomes shorter than the distance to the QPC (about 1~$\mu$m) and the interference fringes are smeared out by thermal averaging \cite{topinka-01-nat}. The conductance is differentiated with respect to tip position in (a) and (b).}
\label{figure-S5}
\end{center}
\end{figure}

%------------------------------------------------------
\section{Phase shift at the ZBA measured in another QPC device}
%------------------------------------------------------

\begin{figure}[h]
\begin{center}
\includegraphics[width=18cm]{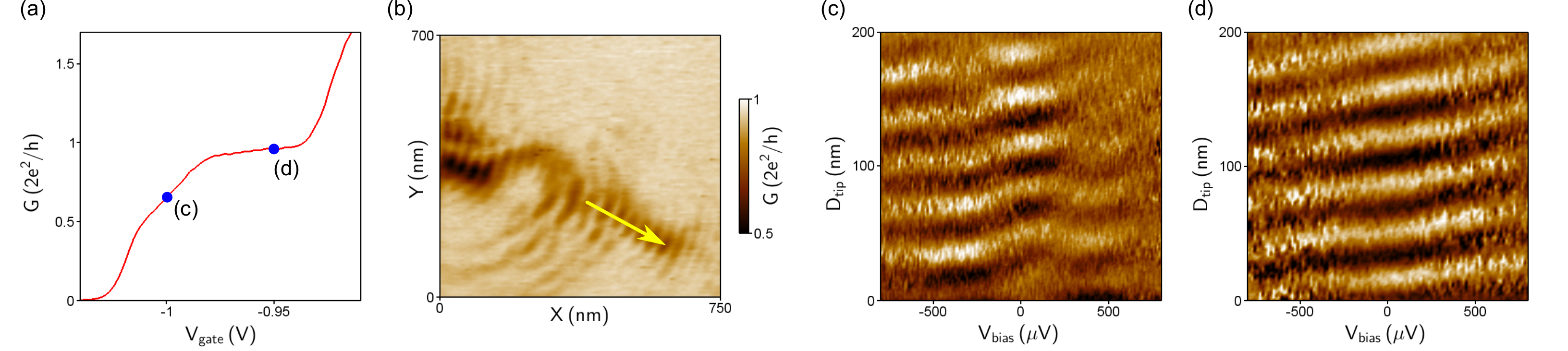}
\caption{Results obtained on a similar QPC device fabricated on the same 2DEG (this sample was studied in Ref.~\cite{brun-14-ncom}). All data are recorded at the base temperature of 20~mK. (a) Conductance versus gate voltage. (b) SGM map at $-0.95$~V gate voltage, $-6$~V tip voltage, and 40~nm tip height. (c,d) Source-drain bias spectroscopy of the interference fringes along the line indicated in (b) at two gate voltages: $-1$~V for (c) and $-0.95$~V for (d). The interference fringes exhibit a phase shift in the bias range of the zero-bias peak in (c) while the fringes are linear on the conductance plateau in (d).}
\label{figure-S6}
\end{center}
\end{figure}

\newpage
%------------------------------------------------------
\section{Model of the scanning gate interferometry experiment}
%------------------------------------------------------

\begin{figure}[b]
\begin{center}
\includegraphics[width=9cm]{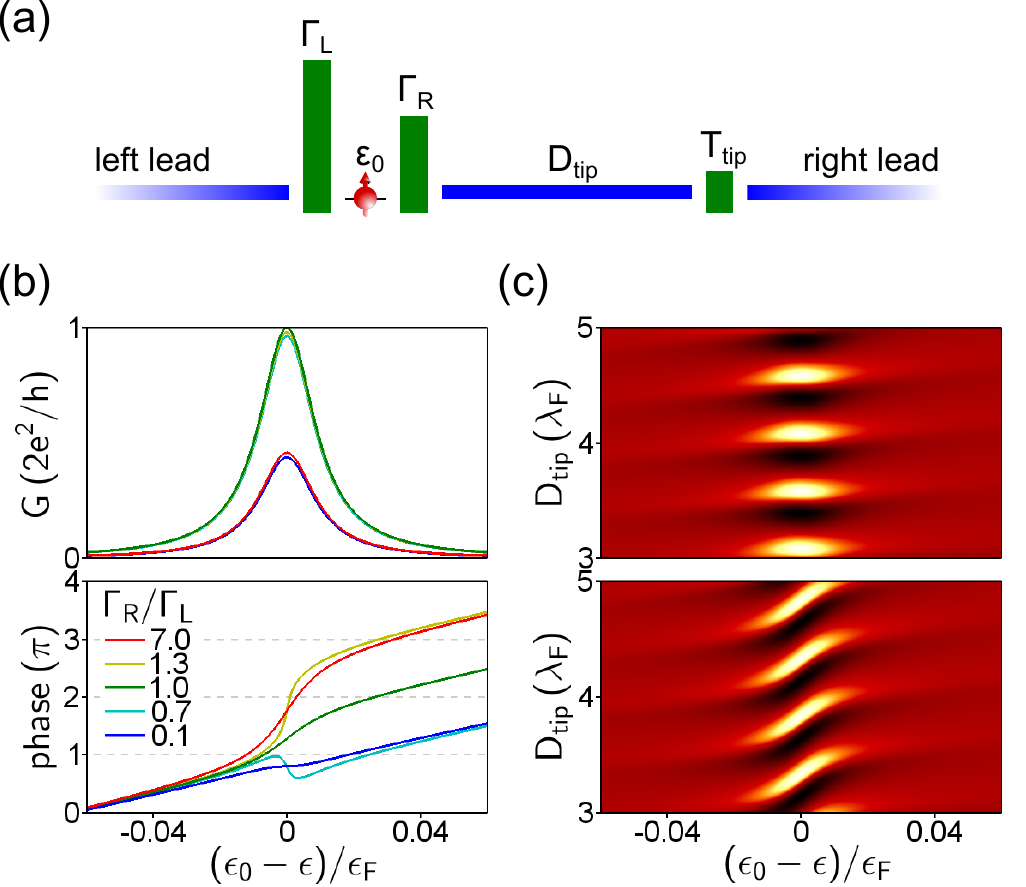}
\caption{(a) Model of the SGM-based interferometry experiment where the QPC is represented by an asymmetric QD with a single level $\epsilon_0$. (b) Conductance of the QD (top) and phase of the tip-induced interference (bottom) as a function of energy $\epsilon$ for different asymmetries of the QD tunneling rates $\Gamma_L$ and $\Gamma_R$. (c) Pattern of tip-induced interference fringes for the blue (top) and red (bottom) curves in (b). The conductance is differentiated with respect to $D_{\rm tip}$ to show only the fringes.}
\label{figure-S7}
\end{center}
\end{figure}

We consider a one-dimensional Fabry-P\'erot (FP) cavity formed between an asymmetric quantum dot (QD), that represents the QPC, and a local potential perturbation, corresponding to the SGM tip (Fig.~\ref{figure-S7}(a)). At zero temperature, the conductance is given by the transmission of the FP cavity:
\begin{eqnarray*}
T_{\rm FP} = \left|\frac{t_{\rm dot}t_{\rm tip}}{1-r_{\rm dot}r_{\rm tip}\exp(2ikD_{\rm tip})}\right|^2
\end{eqnarray*}
where the cavity length $D_{\rm tip}$ is changed by moving the tip, $k$ is the electron wave-vector, $t_{\rm dot},r_{\rm dot}$ and $t_{\rm tip},r_{\rm tip}$ are the complex transmission and reflexion amplitudes of the dot and the tip, respectively. The scattering amplitudes of the dot can be calculated exactly in the non-interacting case, using a tight-binding model made of a single site at energy $\epsilon_0$ connected via two tunnel barriers (hoping terms $V_{\rm L}$ and $V_{\rm R}$) to two semi-infinite leads (hoping term $t$ and lattice parameter $a$):
\begin{eqnarray*}
t_{\rm dot} = \frac{ 2 i \sin(ka) V_{\rm L}V_{\rm R}/t }{ \epsilon(k) - \epsilon_0 + A\cos(ka) + i A\sin(ka) } \\
r_{\rm dot} = - \: \frac{ \epsilon(k) - \epsilon_0 + A\cos(ka) + i B\sin(ka) }{ \epsilon(k) - \epsilon_0 + A\cos(ka) + i A\sin(ka) }
\end{eqnarray*}
where $\epsilon(k)=-2t\cos(ka)$, $A=(V_{\rm L}^2+V_{\rm R}^2)/t$, and $B=(V_{\rm L}^2-V_{\rm R}^2)/t$. Note that the reflection coefficient is for waves arriving from the cavity, i.e. on the right side of the dot. The transmission $T_{\rm FP}$ of the interferometer is calculated for a weak tip back-scattering ($t_{\rm tip}=0.99$) and various asymmetries of the dot barriers $(V_{\rm L}/t,V_{\rm R}/t)=(0.132,0.05)$, (0.109,0.09), (0.1,0.1), (0.093,0.107), (0.0515,0.132). The interference fringes are plotted in Fig.~\ref{figure-S7}(c) as a function of tip distance and electron energy, in two situations corresponding to opposite barrier asymmetries. The simulations are plotted as a function of $-\epsilon$ to be easily compared with the experiments where the conductance is plotted as a function of source-drain bias ($eV_{\rm bias}=-\epsilon$). The global slope of the fringes simply results from the change of the electron wavelength with energy, and the shift around $\epsilon=\epsilon_0$ results from the resonant level in the QD. The phase of the fringes is extracted from the Fourier transform of $T_{\rm FP}$ with respect to $D_{\rm tip}$ and plotted in Fig.~\ref{figure-S7}(b), bottom panel, for the five tunneling rate ratios $\Gamma_{\rm R}/\Gamma_{\rm L}=V_{\rm R}^2/V_{\rm L}^2$ indicated above. Because of the weak tip-induced reflection ($t_{\rm tip}=0.99$), the phase extracted from the interference fringes is close to the reflection phase of the dot:
\begin{eqnarray*}
\phi_{\rm dot} = \arctan\left(\frac{B\sin(ka)}{\epsilon(k)-\epsilon_0+A\cos(ka)}\right) - \arctan\left(\frac{A\sin(ka)}{\epsilon(k)-\epsilon_0+A\cos(ka)}\right) + \pi
\end{eqnarray*}
which was also discussed in Ref.~\cite{buks-96-prl}. In contrast to the transmission $|t_{\rm dot}|^2$ of the dot which follows the well-known Breit-Wigner formula for any barrier asymmetry (Fig.~\ref{figure-S7}(b), top panel), the phase shift that occurs across the resonance strongly depends on the relative values of the tunneling rates (Fig.~\ref{figure-S7}(b), bottom panel). The shift is $2\pi$ when the right barrier is the most transparent because the dot belongs to the cavity (Fig.~\ref{figure-S7}(c), bottom panel), it is only $\pi$ for the symmetric case, and it reduces to zero when the right barrier is the least transparent because the dot is outside the cavity (Fig.~\ref{figure-S7}(c), top panel). As a consequence, the reflection phase measured by scanning gate interferometry is between zero and twice the transmission phase and should be interpreted carefully. This simple model also shows that only smooth phase shifts are expected for non-interacting electrons in contrast to the abrupt phase jumps observed in our experiment. To discuss the theoretical predictions for QDs in the Kondo regime, the solution of the Anderson model with finite Coulomb interaction $U$ in the dot, calculated in Ref.~\cite{gerland-00-prl}, is reported schematically in Fig.~\ref{figure-S8} and discussed below.

%------------------------------------------------------
\section{Expected transmission phase for a Kondo quantum dot}
%------------------------------------------------------

\begin{figure}[h]
\begin{center}
\includegraphics[width=10cm]{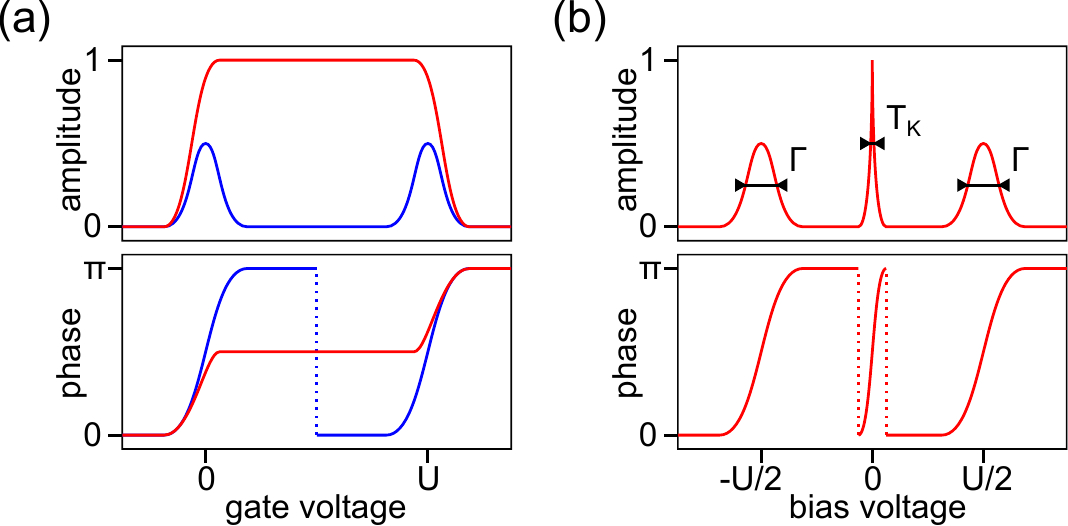}
\caption{Red lines show the amplitude and phase of the transmission coefficient through a QD in the Kondo regime at zero temperature according to Ref.~\cite{gerland-00-prl} versus (a) gate voltage and (b) bias voltage at fixed gate voltage $U/2$. Blue lines in (a) show the Coulomb blockade regime without Kondo correlations.}
\label{figure-S8}
\end{center}
\end{figure}

When a QD is in a Coulomb blocked region with an odd number of electrons, the Kondo effect gives rise to an enhanced transmission amplitude that reaches $2e^2/h$ at low enough temperature and for symmetric couplings to the leads (Fig.~\ref{figure-S8}(a), top panel, red line). The transmission phase of the electrons at low bias is locked at $\pi/2$ in this Kondo valley (Fig.~\ref{figure-S8}(a), bottom panel, red line), according to the Friedel sum rule which relates the phase to the occupation probability per spin (this quantity equals $1/2$ in the Kondo regime due to screening of the unpaired spin by the surrounding conduction electrons). At finite bias voltage, the problem is more complex due to decoherence of the Kondo correlations between the two reservoirs. Assuming that the QD remains in equilibrium, the spectral properties of the transmission coefficient have been discussed in Ref.~\cite{gerland-00-prl}. In the Kondo regime, the transmission amplitude shows a sharp resonance at the Fermi level, whose width is given by the Kondo temperature $T_K$, in addition to the two peaks of width $\Gamma$, separated by the Coulomb charging energy $U$, and corresponding to the spin-degenerate single-particle energy levels (Fig.~\ref{figure-S8}(b), top panel). The transmission phase exhibits a smooth shift from 0 to $\pi$ when the energy is swept across the single-particle energy levels and across the Kondo resonance at zero bias (Fig.~\ref{figure-S8}(b), bottom panel). Between each of these peaks, the phase shows a sharp lapse by $-\pi$, resulting in the ``Kondo double phase lapse'' predicted in Ref.~\cite{gerland-00-prl} at low enough temperature. In absence of Kondo correlations, the phase shows only one lapse by $-\pi$ between the Coulomb blockade peaks corresponding to the spin-degenerate single-particle energy levels (Fig.~\ref{figure-S8}(a), blue lines).

\newpage
%------------------------------------------------------

%------------------------------------------------------

%------------------------------------------------------
\end{document}